\documentclass[graybox]{svmult}

\usepackage{eurosym}

\usepackage[colorlinks=true,linkcolor=black,anchorcolor=black,citecolor=black,filecolor=black,menucolor=black,runcolor=black,urlcolor=black]{hyperref}
\usepackage{amsmath}
\usepackage{amsfonts}  
\usepackage{mathptmx}       
\usepackage{helvet}         
\usepackage{courier}        
\usepackage{type1cm}        

\usepackage{makeidx}         
\usepackage{graphicx}        
\usepackage{multicol}        
\usepackage[bottom]{footmisc}

\makeindex             

\begin{document}

\setcounter{chapter}{2}
\title{Urban Mobility}
\author{Laura Alessandretti and Michael Szell}
\institute{Laura Alessandretti \at DTU Compute, Technical University of Denmark. \email{lauale@dtu.dk}
\and Michael Szell \at NEtwoRks, Data, and Society (NERDS), ITU Copenhagen. \email{misz@itu.dk}}
%
%
\maketitle

\abstract*{In this chapter, we discuss urban mobility from a complexity science perspective. First, we give an overview of the datasets that enable this approach, such as mobile phone records, location-based social network traces, or GPS trajectories from sensors installed on vehicles. We then review the empirical and theoretical understanding of the properties of human movements, including the distribution of travel distances and times, the entropy of trajectories, and the interplay between exploration and exploitation of locations. Next, we explain generative and predictive models of individual mobility, and their limitations due to intrinsic limits of predictability. Finally, we discuss urban transport from a systemic perspective, including system-wide challenges like ridesharing, multimodality, and sustainable transport.}

\abstract{In this chapter, we discuss urban mobility from a complexity science perspective. First, we give an overview of the datasets that enable this approach, such as mobile phone records, location-based social network traces, or GPS trajectories from sensors installed on vehicles. We then review the empirical and theoretical understanding of the properties of human movements, including the distribution of travel distances and times, the entropy of trajectories, and the interplay between exploration and exploitation of locations. Next, we explain generative and predictive models of individual mobility, and their limitations due to intrinsic limits of predictability. Finally, we discuss urban transport from a systemic perspective, including system-wide challenges like ridesharing, multimodality, and sustainable transport.}

\section{Introduction and history of quantitative mobility studies}
\label{sec:empirical}
From commuting to work, running errands, to going on a leisure trip, mobility is an integral part of our daily lives.
Humans allocate a significant amount of time, money and energy to travel, with the average US household spending on mobility a non-negligibly high value of around 16\% \cite{transport_stats}, and with commuting accounting on average for around 56 minutes per day \cite{burd2021travel}.   
Mobility underlies some of the most critical challenges of our times, including the design of sustainable urban transportation and the containment of epidemics. 
For example, a badly designed transport system can drive critical problems such as congestion, road deaths, and pollution. 
Transportation is estimated to account for around $27\%$ of greenhouse gas emissions in the US, out of which around $57\%$ comes from light-duty on-road vehicles (cars, motorbikes, vans) \cite{epa}.  
People on the move drive social mixing which in turn can facilitate the spread of infectious diseases \cite{alessandretti2022human}.
Important societal issues are deeply intertwined with mobility patterns, and it has become critical to quantify and understand human movements. 
Understanding human mobility is important not only from a theoretical standpoint but also crucial to inform policy makers on how to improve urban transport and infrastructure, with the ultimate goal to invest into policies that improve the movement of people or that reduce the need for movement in the first place.

The study of human movements has a long history. 
The first rigorous approaches to quantify human movement patterns can be traced back to the 19th century, for example to Eduard Lill \cite{lill1889grundgesetze}, statistician and chief inspector of the imperial-royal Austrian north-west railway company, who noticed that the number of people traveling from a place decreases as a hyperbola, eventually giving rise to the gravity laws of travel discussed in Chapter~15. 
From this statistical perspective, the aggregate of many people can be described via simple mathematical laws, despite each individual having their individual, unique reasons for movement. 
This perspective is taken to the extreme in statistical physics, where large numbers of particles are easily described via their emergent properties, for example the temperature and pressure of a gas. 
This reductionist Ansatz, and the corresponding tools developed in statistical physics, are powerful ways to understand large numbers of interacting or moving entities \emph{statistically}, whether they are particles or humans \cite{ball2003physical}.

From these early studies on human mobility, we have come a long way. 
As an empirical science, the availability of high-resolution data has been crucial to drive evidence-based understanding. 
Such data have started to become available at the start of the 21st century, with first the diffusion of mobile phones, smartcards, and GPS positioning systems installed on vehicles, then smartphones and wearable devices. 
Concurrently with the development of increasingly powerful computational tools, the data revolution has enabled a richer and richer understanding of human mobility patterns.
First, scientists could shift focus from the collective to the individual level, by studying single trajectories that have increasingly high temporal and spatial resolution, to the point that we can now trace actual route choices. 
Second, we are achieving a better and better understanding of the mechanisms underlying mobility behavior, through enriching mobility data with additional information, such as the features of the visited locations, individuals' social interactions, environmental and weather data, and features of the built environment \cite{resch2019hds}.

In this chapter, we give an overview on how the field of complex systems, shaped considerably by statistical physicists, makes use of and adapts their tools in the context of urban mobility. 
Our chapter does not claim completeness but is an introductory overview of the fundamental concepts with hand-picked highlights; for a comprehensive review see Ref.~\cite{barbosa2018human}. 
We outline the main aspects of the field by following largely the historical co-evolution of data and research. 
In Section~\ref{sec:data} we discuss sources of data and issues of data quality.
In Section~\ref{sec:empirical} we introduce metrics and descriptive results about the statistical properties of individual empirical trajectories. 
In Section~\ref{sec:modeling} we summarize the main statistical and mechanistic models about urban mobility patterns. 
Although we explicitly do not cover other approaches to urban mobility, for example from transport engineering, we follow up with an outlook in Section~\ref{sec:transportation} on sustainable transport and urban livability due to the urgency of this topic in the context of the climate crisis, asking to which extent some of the long-range or high-speed movements in cities of today are actually necessary and how they could be avoided. 
Finally, we provide a list of important tools and further materials in Section~\ref{sec:tools} and end with conclusions in Section~\ref{sec:conclusion}.

\section{Data sources \label{sec:data}}
Since the early 2000s, the development and widespread adoption of technologies that record the positions of individuals over time has driven a rapid growth of the field of human mobility, through enabling access to data \cite{resch2019hds}. Here, we describe some of the most widely used data sources. 

\textbf{Data from mobile network operators.} Mobile network operators gather location data for billing and operational purposes, through monitoring the cell towers that user devices connect to. These data have been widely used for human mobility research. 
Data collected by network operators has inherently limited spatial resolution, because the coverage of cell towers is quite wide, ranging from some tens of meters in urban areas up to tens of kilometers in rural areas. 
In terms of temporal resolution, there has been a rapid evolution over the last 20 years. 
Before the diffusion of smartphones, research was largely based on Call Detail Records (CDR), which captures the closest cell tower to individuals as they issue or receive calls and SMS. 
More recently, researchers have gained access to higher-frequency data. 
eXtended Detail Records (XDR) are collected when users explicitly request an http address or when the phone downloads content from the Internet (e.g., emails, messages, app updates).  
Control Plane Records (CPRs) are network-triggered (e.g., assigning a new antenna, connecting new devices) and are used to monitor the cellphone network status. 
CDR, XDR, and CPR vary signiﬁcantly in their time granularity and data sparsity \cite{pappalardo2021evaluation}. 

\textbf{Data from smartphone applications.} Other useful data sources are represented by services that collect GPS positions through smartphone applications or services. 
In the early years of human mobility research, these data would mostly be collected by Location Based Social Networks (LSBN) services, such as Foursquare or Twitter, which would gather location data only when individuals used the service. 
Nowadays, other applications, including navigation apps or apps that perform targeted advertising, collect GPS positions from user devices at higher frequency. 
In turn, these applications are often equipped with software development kits (SDK) that can send the GPS data from the user’s device to the company that produced the SDK. 
Through using several smartphone applications, SDK companies can thus collect comprehensive geolocation data.
Especially since the COVID-19 pandemic, SDK data has become widely used for mobility research \cite{aleta2020modelling,chang2021mobility}.

\textbf{Transport and mobile sensor data.} The digitization of private and public transport services now allows tracking of citizens in the public transportation system, such as through public transport travel card, and analyzing/visualizing entire taxi systems and transportation fleets. Further,
detailed records are being generated by novel mobility sharing systems, from car and bicycle sharing to e-stroller and ride sharing. Custom sensors, installed on vehicles, can provide the potential to sense ecological urban variables and the sentiments of city dwellers in unprecedented detail. 

Most data sources come with a wide range of biases and measurement problems, such as missing (sub)trajectories, zigzagging, or GPS jitter. Mobility data usually originate from applications that were not developed for research purposes, leading to typically low data quality. Any serious movement data analysis therefore has to account for adequate data preprocessing, including assessment and mitigation of data quality issues, which is often an extensive task \cite{andrienko2016understanding}. Further, stop and home detection are not trivial \cite{calabrese2013understanding,JSSv103i04}.

\section{Statistical properties of individual trajectories}
\label{sec:empirical}

Characterizing the statistical properties of individual trajectories is necessary to understand the underlying dynamics of human mobility and to design reliable predictive models.
Generally, one can think of the movements of an individual as a trajectory consisting of \emph{displacements} between locations and \emph{pauses} at locations where the individual stops and spends time. 
Depending on the sampling frequency, the availability of spatio-temporal details, and the goal or scope of the analysis, trajectories can be studied either as sequences of locations, as sequences of locations embedded in space and time, or as high-frequency trajectories. 
In this section, we review some of the fundamental metrics that have been used in the literature to characterize individual trajectories and the resulting empirical findings. 
We will introduce richer and richer representations of individual trajectories. 
The entropy measures that we introduce here swiftly can be well motivated from statistical mechanics, and can be extended or applied to many additional questions in urban science, see Chapter~5.

\subsection{Trajectories as sequences of locations}
In the simplest form, a trajectory can be represented as a sequence of locations, discarding aspects related to the position of locations in physical space. A large stream of literature has focused on the characteristics of these sequences of places. 

One first key question is: How predictable are these sequences of locations? Or, in other words, to what extent is human mobility repetitive? This question is important for modeling purposes, to understand how simple we can make our models, and for privacy considerations. A key metric to address these questions is entropy, which can be thought of as a measure of uncertainty.

The most naive way of assessing predictability of an individual $i$ is by examining their visited number of unique locations $N_i$. In the trivial case $N_i=1$ there is no uncertainty about the individual's location and it is fully predictable. However, the larger $N_i$ gets, the more uncertain the location and the less predictable the individual becomes. A measure for the uncertainty of an individual's location which considers only $N_i$ was introduced by Ref.~\cite{song2010modelling} as the random entropy:
\begin{definition}[Random entropy]
The random entropy $S^{\mathrm{rand}}_i$ of an individual $i$'s trajectory, given their visited number of unique locations $N_i$, is
\begin{equation}
    S^{\mathrm{rand}}_i = \log_2 N_i. 
\end{equation}
\end{definition}

\begin{svgraybox}
Example: Person 1 visited 2 locations. Therefore, $S^{\mathrm{rand}}_1 = \log_2 N_1 = \log_2 2 = 1$.
Example: Person 2 visited 4 locations. Therefore, $S^{\mathrm{rand}}_2 = \log_2 N_2 = \log_2 4 = 2$.
\end{svgraybox}

Coming from thermodynamics and statistical physics, entropy more formally measures the degree to which the probability of a system is spread out over different possible microstates. In information theory it is called Shannon entropy:
\begin{definition}[Shannon entropy, temporal-uncorrelated entropy]
The Shannon entropy $H(X)$ of a discrete random variable $X$ is
\begin{equation}
    H(X) = -\sum p(x)\log_2 p(x), 
\end{equation}
where $\sum$ denotes the sum over all possible values $x$ of $X$. In mobility research, the Shannon entropy of an individual $i$'s trajectory has been called ``temporal-uncorrelated entropy'' \cite{song2010modelling}
\begin{equation}
    S^{\mathrm{unc}}_i = -\sum_{j=1}^{N_i} p_i(j)\log_2 p_i(j), 
\end{equation}
where $p_i(j)$ denotes the fraction of individual $i$'s visits to location $j$.
\end{definition}

\clearpage
\begin{svgraybox}
Example: Person 1 visited location A 270 times and location B 30 times. Then $p_1(A) = 0.9$, $p_1(B) = 0.1$, and
\begin{align*}
  S^{\mathrm{unc}}_1 &= -\sum_{j=1}^{N_1} p_1(j)\log_2 p_1(j) \\
   &= -\bigl( p_1(A)\log_2 p_1(A) + p_1(B)\log_2 p_1(B)\bigl)\\
   &= -\bigl( 0.9\log_2 0.9 + 0.1\log_2 0.1\bigl)\\
   &\approx 0.47
\end{align*}
Example: Person 2 visited location A 200 times, location B 300 times, location C 300 times, location D 200 times, meaning $p_2(A) = 0.2$, $p_2(B) = 0.3$, $p_2(C) = 0.2$, and $p_2(D) = 0.2$. Then $S^{\mathrm{unc}}_2 \approx 1.97$.
\end{svgraybox}

The random and temporal-uncorrelated entropies of an individual are identical, $S^{\mathrm{rand}}=S^{\mathrm{unc}}$, when all locations are visited with equal probability. However, $S^{\mathrm{unc}}$ becomes smaller the more unequal, or skewed, the distribution probabilities of different locations are. From the examples above, $S^{\mathrm{rand}}_2 = 2$ which is almost the same as $S^{\mathrm{unc}}_2 = 1.97$ because all fours locations were visited with similar frequency. However, for individual 1, $S^{\mathrm{rand}}_1 = 1$ is much larger than $S^{\mathrm{unc}}_1 = 0.47$ because in this case the visitation frequencies were much more skewed, 270 to 30. These two concepts of entropy, random and temporal-uncorrelated, are thus useful to assess how much the skew in visiting probabilities affects the uncertainty of locations.

Extending the concept of entropy one step further adds the aspect of ordering, asking: How heterogeneous are visitations not only across locations but also in their time-ordering?
\begin{definition}[Real entropy]
The ``real'' entropy $S^{\mathrm{real}}_i$ of an individual $i$'s trajectory is given by
\begin{equation}
    S^{\mathrm{real}}_i = -\sum_{T'_i\subset T_i} P(T'_i)\log_2 P(T'_i)
\end{equation}
where $P(T'_i)$ is the probability of finding a particular
time-ordered subsequence $T'_i$ in the trajectory $T_i$ \cite{song2010modelling}.
\end{definition}
Thus, an individual with a trajectory of repeating or regular patterns, such as \verb%ABACDCABACDCABACDC% will have a much lower real entropy than an individual who has an identical number and visitation distribution of locations but a less regular order of visitations, for example \verb%ABCDACABCDCABACACD%. Similarly as $S^{\mathrm{rand}}$ was an upper limit for $S^{\mathrm{unc}}$, here $S^{\mathrm{unc}}$ is an upper limit for $S^{\mathrm{real}}$. The identity $S^{\mathrm{unc}} = S^{\mathrm{real}}$ is given when the order of locations in the trajectory is randomized, thereby destroying all time-correlated information.

A measure associated to the entropy $S$ is the predictability $\Pi$, derived from information theory \cite{song2010modelling}, which is the probability that an optimal predictive algorithm can predict correctly the individual’s future whereabouts. Every type of entropy has a respective predictability measure, for which the inequality relation between entropies $S^{\mathrm{rand}} \geq S^{\mathrm{unc}} \geq S^{\mathrm{real}}$ is inverted: $\Pi^{\mathrm{rand}} \leq \Pi^{\mathrm{unc}} \leq \Pi^{\mathrm{real}}$. The more uncertainty, the less predictability. As Ref.~\cite{song2010modelling} have shown, a significant share of predictability is encoded in the time ordering which make the trips of most individuals highly predictable.

This high predictability of human trajectories has strong privacy implications. For example, it was shown that for the trajectories of 1.5 million individuals, with hourly sampling and spatial aggregation to 6500 cell phone towers, only 4 random spatio-temporal points are needed to identify 95\% of individuals \cite{de2013unique}. Such insights have profound legal and ethical consequences on the sharing and tracking of micro-mobility data, and have spurred research on privacy-enhancement techniques such as cloaking, suppression, aggregation, swapping, or differential privacy \cite{fiore2020privacy}.

While the statistical characterization of trajectories, either individually or in aggregate, is crucial for fundamental predictability and privacy considerations, two important practical questions remain: 1) What are those regular sequences of locations, and 2) What is the relation with other people's movements in the city? The first question can be tackled via motif analysis, where the most common trips of individuals are interpreted as a network of trips between locations \cite{schneider2013unravelling}. Motifs in complex networks are generally small subgraphs that appear with significantly higher than expected probability \cite{milo2002network}. Analyzing both surveys and mobile phone data in different cities, Ref.~\cite{schneider2013unravelling} found that half of the daily mobility networks are just described by two trivial motifs, consisting of one node (no movement) or of two nodes (back and forth movement, typically home-work). The vast majority of all other daily movements are described by remaining 15 motifs. The second question -- the relation with other people's movements -- has been investigated for example in the context of bus transit patterns of 5 million individuals in Singapore \cite{sun2013understanding}. Due to the striking regularity of daily commutes, the emergence of ``familiar strangers'' in a tight-knit contact network was observed, with crucial consequences on the impact of human behavior on diffusion/spreading processes. More recent studies have compared explicitly social relations with mobility behavior \cite{sekara2016fundamental,chen2022contrasting}. For example, the Copenhagen network study \cite{sekara2016fundamental} has found striking differences in geospatial versus social entropy in a population of 1000 university students, for example uncovering ``party nights'' that are characterized by geospatial exploration but conservative social behavior.

The reason why sequences of human locations are highly predictable is rooted in the high level of regularity of day-to-day routines, with mobility highly dominated by home-work commuting. Moreover, the way humans visit places and allocate time among them is characterized by universal properties, as shown by several empirical studies. Some notable properties include: 

\emph{Heterogeneous distribution of visits.} Humans allocate their time in a heterogeneous way across places, implying they spend most of their time within a small set of favorite locations. Ranking locations from the most visited to the least visited, the number of visits for a location with rank $L$ follows a power-law $P \sim L^{-\alpha}$ with exponent $\alpha\sim1.2$ \cite{song2010modelling, alessandretti2018evidence}.

\emph{Sublinear exploration.} The number of distinct locations $S$ visited by an individual grows over time (t) as $S\sim^{\beta t}$, with $\beta\sim0.6$, implying that there is a decreasing tendency to explore new locations \cite{song2010modelling}.

\emph{Periodicity.} The probability to return to a location visited n hours before is characterized by peaks at 24h, 48h and 72h, due to the recurrence and temporal periodicity inherent to human mobility, which is driven by circadian and weekly patterns. \cite{song2010modelling}

\emph{Conservation of the number of visited places.} It was shown that, despite the fact that our routines change and evolve over long time-scales, some properties of the set of visited locations, including their size remain approximately conserved at the level of the single individual \cite{alessandretti2018evidence}. 

\subsection{Trajectories in space and time}
If the data are available, realism can be increased by considering trajectories as sequences of locations embedded in space and time. Two key properties that characterize these sequences are the waiting time and the displacement distributions. 

\begin{definition}[Waiting time distribution]
The waiting time distribution $P(\Delta t)$ describes probabilistically the time $\Delta t$ that an individual pauses after a move.
\end{definition}
\begin{definition}[Displacement distribution]
The displacement distribution $P(\Delta r)$ describes probabilistically the distance $\Delta r$ that an individual moves after a pause.
\end{definition}

The distribution of waiting times (or pause durations), $\Delta t$, between movements and the distribution of distances, $\Delta r$, travelled between pauses are useful to quantitatively assess the dynamics of human mobility. For example, specific probability distributions of distances and waiting times characterise different types of diffusion processes.

Thanks to the recent availability of data used as proxy for human trajectories, the characteristic distributions of distances and waiting times between consecutive locations have been widely investigated. There is no agreement, however, on which distribution best describes these empirical datasets. Pioneer studies, based on CDR~\cite{song2010modelling, gonzalez2008understanding} and banknote records~\cite{brockmann2006scaling}, found that the distribution of displacement $\Delta r$ is well approximated by a power-law, $P(\Delta r) \sim \Delta r ^{-\beta}$, (or ``L\'evy distribution''\cite{baronchelli2013levy}, typically with $1<\beta < 3$), and that an exponential cut-off in the distribution may control boundary effects \cite{gonzalez2008understanding}. These findings were confirmed by studies based on GPS trajectories of individuals~\cite{wang2014correlations, zhao2015explaining,rhee2011levy} and vehicles~\cite{jiang2009characterizing,liu2012understanding}, as well as online social networks data~\cite{beiro2016predicting,cheng2011exploring,hawelka2014geo}. It has been noted, however, that power-law behaviour may fail to describe intra-urban displacements~\cite{noulas2012tale}. Other analyses, based on online social network data~\cite{wu2014intra,liu2014uncovering, jurdak2015understanding} and GPS trajectories~\cite{liu2015crossover,liang2012scaling,gong2016inferring,zhao2015automatic} showed that the distribution of displacements is well fitted by an exponential curve, $P(\Delta r) \sim e^{-\lambda \Delta r}$, in particular at short distances. Finally, analyses based on Taxi GPS~\cite{wang2015comparative,tang2015uncovering} suggested that displacements may also obey log-normal distributions, $P(\Delta r) \sim (1/\Delta r) * e^{-(\log \Delta r - \mu)^2/2 \sigma^2} $. In Another study found that this is the case also for single-transportation trips \cite{zhao2015explaining}.

Fewer studies have explored the distribution of waiting times between displacements, $\Delta t$, as trajectory sampling is often uneven; e.g., in CDR data location is recorded only when the phone user makes a call or sends an SMS, and LBSN data include the positions of individuals who actively ``check-in'' at specific places. Analyses based on evenly sampled trajectories from mobile phone call records~\cite{song2010modelling,schneider2013unravelling}, and individuals GPS trajectories~\cite{wang2014correlations,rhee2011levy} found that the distribution of waiting times can also be approximated by a power-law.
A recent study based on GPS trajectories of vehicles, however, suggests that for waiting times larger than $4$ hours, this distribution is best approximated by a log-normal function~\cite{gallotti2016stochastic}. Several studies have highlighted the presence of natural temporal scales in individual routines: distributions of waiting times display peaks that correspond to the typical times spent home on a typical day (around 14~hours) and at work (3-4~hours for a part-time job and 8-9~hours for a full-time job) \cite{schneider2013unravelling,hasan2013spatiotemporal,bazzani2010statistical}. 

The datasets considered have different \emph{spatial resolution and coverage}, and few studies have so far considered the whole range of displacements occurring between $10^1\,\mathrm{m}$ and $10^7\,\mathrm{m}$ ($10,000~\mathrm{km}$). 
Another issue for comparability concerns the \emph{temporal sampling} in the datasets analysed so far. Uneven sampling typical of CDR and LBSN data (i) does not allow to distinguish phases of \emph{displacement} and \emph{pause}, since individuals could be active also while transiting between locations, and (ii) may fail to capture patterns other than regular ones~\cite{ccolak2015analyzing,ranjan2012call}, because individuals' voice-call/SMS/data activity may be higher in certain preferred locations.
Finally, studies focusing on displacements effectuated using one or several \emph{specific transportation modalities} (private car~\cite{gallotti2016stochastic,gallotti2015understanding}, taxi~\cite{zhao2015automatic}, public transportation~\cite{roth2011structure}, or walking~\cite{rhee2011levy}) capture only a specific aspect of human mobility behaviour. 

Nevertheless, an aggregate measure of diffusion over many individuals, which describes how fast an area is explored, is given by the mean square displacement:

\begin{definition}[Mean square displacement]
The mean square displacement $\mathrm{MSD}(t)$ measures the deviation of the position of $N$ individuals from a reference position over a timespan $t$ as an (ensemble) average over all the individuals $j$:
\begin{equation}
    \mathrm{MSD}(t) = \langle \mathbf{r}_j(t)-\mathbf{r}_j(0) \rangle = \frac{1}{N}\sum_{j=1}^{N}\left(\mathbf{r}_j(t)-\mathbf{r}_j(0)\right)^2,
\end{equation}
where $\mathbf{r}_j(t)$ is the position of individual $j$ at time $t$ and $\mathbf{r}_j(0)$ is their initial reference position.
\end{definition}
The mean squared displacement is a measure introduced in statistical mechanics to study diffusion processes, and can be thought of as measuring the portion of the system ``explored'' by an ensemble of individuals. 
Empirical research has shown that human mobility patterns follows a slow diffusion process, in which the MSD grows logarithmically~\cite{song2010modelling}.

To distinguish whether individuals all sample from the same displacement or waiting time distributions, or whether individuals have different movement behaviors, it is useful to define the radius of gyration:
\begin{definition}[Radius of gyration]
The radius of gyration $r_g$ is the typical distance travelled by an individual over a fixed timespan $t$,
\begin{equation}
    r_g = \sqrt{\frac{1}{n}\sum_{i=1}^{n} |\mathbf{r}_i-\mathbf{r}_{\mathrm{cm}}|^2 }, 
\end{equation}
where $n$ is the number of jumps within timespan $t$, $\mathbf{r}_i$ represents the position at step $i$, and $\mathbf{r}_{\mathrm{cm}} = \frac{1}{n}\sum_{i=1}^{n} \mathbf{r}_i$ is the center of mass of the trajectory.
\end{definition}

Ref.~\cite{gonzalez2008understanding} have introduced this metric to human mobility studies, showing that individuals have different typical travel distances.
In particular, it has been shown that the distribution of the radius of gyration across individuals is broad and can be well described by a truncated power-law~\cite{gonzalez2008understanding}.

\section{Modeling urban mobility patterns}
\label{sec:modeling}
To study the movements of humans, the field of complex systems adapts models from statistical physics, which studies the movements of particles. Central to this approach is the concept of random walk. A random walk describes the random, Brownian motion of particles suspended in a liquid or gas, and can be taken as the simplest model for a moving individual. 

\begin{definition}[Random walk]
A random walk is a random process describing a path of random steps on a mathematical space.
\end{definition}

\begin{svgraybox}
Example: Random walk on the integer number line $\mathbb{Z}$ which starts at 0 and at each step moves $-1$ or $+1$ with equal probability, for example decided by a coin flip. Thus, after two coin flips, the following outcomes are possible: 
\begin{itemize}
    \item $-2$, after $0-1-1$
    \item \phantom{$-$}0, after $0-1+1$ or $0+1-1$
    \item \phantom{$-$}2, after $0+1+1$
\end{itemize}
\end{svgraybox}

In the above example, the random walk describes a trajectory on $\mathbb{Z}$ with a fixed jump distance of 1 and an arbitrary waiting time -- for repeated coin flipping this could be for example 3 seconds. Further, this random walk is memoryless, as the coin flip is a random process independent of previous outcomes. There are several ways to extend the behavior of random walkers, adding complexity to more realistically model human mobility. For example, in continuous-time random walk (CTRW) models, jump lengths and waiting times can follow arbitrary distributions. This model was suggested as one of the first models able to capture the probability distributions of displacements and waiting times, where the step-lengths and waiting times were calibrated from heavy-tailed probability distributions observed in human mobility \cite{brockmann2006scaling}.

However, human mobility is not completely random, and researchers quickly realized that more realistic mechanisms were needed to capture the properties of human trajectories. An ingredient to make trajectories more realistic is memory, i.e.~individuals can remember and be influenced by visits to previous places. In 2010, Song et al.~\cite{song2010modelling} proposed an agent-based model with memory that could capture many of the observed properties of mobility.

\begin{definition}[EPR model]
 The Exploration and Preferential Return model (EPR) proposes that whenever individuals visit a location, they either (i) explore with probability $P_{\mathrm{new}} = S^{-\gamma}$, where S is the number of distinct locations previously visited, or (ii) return to a previously visited location with complimentary probability $P_{\mathrm{ret}}=1-P_{\mathrm{new}}$. When returning to a previously visited location, the probability $p_i$ to visit location $i$ is proportional to the number of visits the individual previously had to that location, an effect known as preferential return.
\end{definition}

Several variations of the EPR model have been proposed to account for other aspects of human mobility including more memory and recency effects ~\cite{alessandretti2018evidence,szell2012understanding,barbosa2015effect}, social interactions~\cite{schlapfer2021universal}, and heterogeneities across individuals \cite{pappalardo2015returners}. 

An important development of the EPR model is the Preferential Exploration and Preferential Return (PEPR) model, developed by Schläpfer et al. \cite{schlapfer2021universal}. 
\begin{definition}[PEPR model]
The PEPR model couples the movements of agents, so that, when exploring new locations, they are preferentially attracted towards highly frequented areas.
\end{definition}

Importantly, this model is able not only to reproduce the individual mobility, but also collective patterns of mobility, thus linking the literature streams on individual and collective mobility (collective mobility is covered in Chapter 14 of this book). 
In particular, the PEPR model accounts for an important empirical discovery made by Schläpfer et al.~\cite{schlapfer2021universal}, related to the frequency at which individuals visit different locations. 
By studying mobility traces extracted from phones the authors had revealed that the number of users who visit a location at distance $r$, exactly $n$ times in a period of length $T$, decreases as $N\sim r^{-2}f^{-2}$, where $f = n/T$.

While capturing many properties of human movements, preferential return models are still unable to explain the power-law distribution of displacements in mobility data. 
The observation that mobility patterns are captured by power-law distributions seemed to show that human mobility is scale-free in a truly fundamental sense. 
But concurrently, it was known that the natural and built environment is rich in spatial scales, from neighbourhoods, to cities, regions, countries and continents \cite{pumain2006hierarchy}. 
The notion of scale is fundamental within the fields of geography and spatial cognition, because we think of space in a hierarchical fashion characterized by typical scales \cite{grauwin2017ims}.
In this sense, there appeared to be an important schism between traditional geography and the data-driven work on human mobility. 
This divide got explicitly addressed in 2020 by Ref.~\cite{alessandretti2020scales} working with a very large dataset of GPS traces. 
This study showed that typical scales are indeed embedded within the same type of scale-free trajectories analyzed by previous researchers.
The explanation is that scales manifest as containers of mobility behavior.
Neighborhoods have typical sizes, but the distance between the neighborhoods a certain individual visits is unrelated to the size of neighborhoods (it is rather set by the city scale). 
Similarly with cities which also tend to have a typical geographical size, but the cities a certain individual tends to visit may be close – or located in opposite ends of their home country. 
A similar logic is true for scales across neighborhoods, cities, regions, countries, continents, and the entire world. 
The authors use a model-based approach to infer the typical sizes of these containers across millions of individuals.

\begin{definition}[Container model]
Physical space is modelled as a hierarchy of $L$ levels, ordered from the smallest to largest (e.g.~individual locations to countries). 
At any level $l$, space is partitioned into compact containers with a characteristic size. 
Each geographical location $k$ can be identified as a sequence of nested containers, $k  = ( k_1,..., k_l, ..., k_L)$ that contains it. 
The \emph{level-distance} $d(j,k)$ between locations $j$ and $k$ is defined as the highest index at which the two sequences of containers describing $j$ and $k$ differ.
For an agent located in $j$, the probability of moving to $k$ is the product of two factors: $ P( j \rightarrow k)= p_{d(j,k), d(j,h)} \prod _{l \leq d(j,k)}a(k_l)$. 
The first factor, $p_{d(j,k)}$, is the probability of traveling at level-distance $d(j,k)$. 
The second factor $ \prod _{l \leq d(j,k)}a(k_l)$ is the probability of choosing a specific location $k$ at that level-distance, where $a(k_l)$ is the attractiveness of a container at level $l$ including location $k$.
\end{definition}

\section{Improving urban transport systems}\label{sec:transportation}

The research described so far focuses mostly on understanding where and when people move, statistically describing individual trajectories or emerging aggregate patterns. The question of how people get to their destinations -- which transport mode they use, which route they take and why, and the role of the underlying transport system -- has remained largely uncharted from a quantitative perspective, also due to the limited availability of data with high enough resolution. For example, how individuals choose routes through balancing the interplay between public and private forms of transportation such as walking, driving and cycling remains poorly characterized. The existing empirical research using a \emph{systemic} and \emph{long-term} view on urban transport systems, including shared travel, multimodal travel, or system dynamics like induced demand is thus quite limited~\cite{alessandretti2022mum}. In this section we review first on the one hand some microscopic and mode-specific wayfinding mechanisms, on the other hand more systems-level approaches of how to think about the dynamics of urban transport systems with the aim to improve them. We discuss the importance of different urban transport modes, to help design more efficient and sustainable urban transportation. 

\subsection{Wayfinding}
Apart from a statistical, quantitative description of human mobility, it is important to understand the mechanisms of how humans are navigating in cities to find their ways. While pathfinding on a graph (such as a street network) is a solved problem in computer science via Dijkstra's shortest path algorithm and its modern replacements \cite{delling2009engineering}, the implementation of routing on multimodal networks can be challenging \cite{poletti2017public,li2012multimodal}, and an array of cognitive peculiarities have to be accounted for such as upbringing \cite{coutrot2022entropy}, vector-based navigation \cite{bongiorno2021vector}, information overload \cite{gallotti2016lost}, or mode-specific travel behavior \cite{broach2012cyclists}. Recent findings have established that pedestrians, cyclists, and motorists do not tend to follow optimal routes, and are instead influenced by urban features and goal distance and direction \cite{manley_shortest_2015, malleson_characteristics_2018, bongiorno2021vector, lima2016understanding}.

\subsection{Shareability networks}
Existing urban transport systems are highly inefficient, such as the individualized transport mode of taxis. In New York city, for example, a large fleet of around 13,500 taxicabs serves individualized mobility needs, but with a high fraction of idle or low occupation runs. To quantify how much such a localized fleet could be improved, Ref.~\cite{santi2014qbv} introduced the concept of shareability network in 2014 using methods from network science.

The idea of shareability networks is to view a vehicle fleet from a global perspective and to optimize for the whole system with global knowledge instead of locally. For example, instead of having two taxis deliver two individuals in parallel from the same starting point to the same end point, the routes of the two passengers could have been bundled with just one taxi with small inconvenience to both. Taking this idea to the extreme, towards $k$-sharing with $k>2$, would lead to a system of dynamically routed ``taxi buses''. As Ref.~\cite{santi2014qbv} showed, in New York City indeed a large fraction of trips could be shared with relatively low discomfort, with cumulative trip length cut by $40\%$ or more. 

Trip-sharing systems were introduced later by UBER (called UBERPool), and by similar ride hailing companies, leading to a claimed short-term reduction of driven kilometers \cite{uberEarthSunday}. However, in total, companies like UBER have later shown to have generated more traffic and congestion due to rebound effects and their competition with public transport \cite{henao2019impact,tirachini2020ride,diao2021impacts}. This example shows how long-term systems thinking is necessary to understand the full impact and possible unintended consequences of short-term optimization, see Section~\ref{subsec:systemsthinking}.

The research on shareability networks was later expanded with a focus on fleet size and vehicle-sharing networks \cite{vazifeh2018addressing}, finding a potential to reduce the New York city taxi fleet size by 30 percent. Such a potential reduction follows directly from a reorganization of taxi dispatching that could be implemented with a mobile phone app and does not assume ride sharing, nor requires changes to regulations, business models, or human attitudes towards mobility to become effective. Finally, further research uncovered scaling effects in ride-sharing \cite{tachet2017slu,molkenthin2020scaling}.

\subsection{Sustainable, multimodal transport and systems thinking \label{subsec:systemsthinking}}
Understanding urban mobility is impossible without understanding its history and the complex socio-technical system it is part of. This history shows a transition towards car-centricity in the 20th century for most cities, entailing critical consequences on the whole system, for example: over 1.3~Mio.~people die on the road every year \cite{who2018gsr}, vehicular pollution causes millions of more yearly deaths \cite{caiazzo2013air}, and vehicular traffic noise has shown to cause dementia-related diseases on a large scale \cite{cantuaria2021residential}. Further, cars come with massive inefficiencies due to their skewed space requirements and usage patterns \cite{szell2018cqv}. Apart from empirical evidence that this is unsustainable \cite{banister2005unsustainable}, also mathematical models from complexity science show that cities as (car-centric) transport monocultures are not sustainable \cite{louf2013modeling, prietocuriel2021pte}, and that a mere replacement of fossil fuel cars with electric vehicles is not an adequate solution \cite{creutzig2015trc,milovanoff2020elv,brand2021ccm,henderson2020evs}, especially when time-tested, much more efficient and economic, solutions such as mass transit or bicycles are available.

To fix such monocultures, the biggest question in urban transport is how to reverse car dependency \cite{itf2021rcp}, i.e. how to best replace unsustainable modes of transport and to prioritize sustainable ones. In particular, active travel such as walking and cycling has the highest societal benefits \cite{who2022wcl}: Cost-benefit analysis which accounts for the environment and public health reveals that each kilometer walked or cycled in the European Union provides \euro\,0.37 or \euro\,0.18 to society, respectively, while each kilometer driven by car incurs a cost of \euro\,0.11 \cite{gossling2019social}. Overcoming transportation monocultures is also possible with a focus on more multimodal transport, i.e.~the combination of multiple transport modes promising all of their benefits  while avoiding their weaknesses. For an overview of state of the art complex systems based approaches to multimodal mobility see Ref.~\cite{alessandretti2022mum}.

Systems thinking is the appropriate method of understanding how (un)sustainable urban mobility emerges over time \cite{oecd2021tsn}: There is a complex co-evolution between urban transport infrastructure, land use, and socio-political/cultural factors with long-term feedback loops. Complex systems dynamics are crucially dependent on the \emph{interactions} between its parts and not just on the parts themselves. 
\begin{svgraybox}
Example: Increased traffic volume creates pressure to invest into road infrastructure, which drives more traffic volume. This is the well-known phenomenon of \emph{induced demand}. Due to increased roadway capacities, catchment areas increase, leading to urban sprawl, which in turn increases traffic volume and erodes active modes of transport due to increased distances and lack of adequate transport infrastructure. 
\end{svgraybox}
The general phenomenon of car-dependence can be thought of as a dynamic reinforced through different socio-technical systems such as the automotive industry, land use, or car culture \cite{mattioli2020political}. Therefore, long-term urban planning cannot detach transport planning from the planning of land-use, housing, parking \cite{shoup2021high}, etc., but the whole system must be understood as being shaped by strong, non-linear, long-term interactions. Although the opposite of systems thinking, namely short-term engineering thinking, has its use cases, it is inadequate to apply to most problems in complex systems, because the optimization or control of sub-systems can lead to unintended consequences (such as induced demand or urban sprawl) \cite{oecd2021tsn}. Using short-term traffic engineering thinking has lead to the outdated ``predict and provide'' logic, where e.g. traffic volumes are predicted from past flows and the role of policy makers is to react by providing a short-term solution. On the contrary, following systems thinking, this logic is replaced with the more adequate ``decide and provide'' principle, which aims to pro-actively shape the future \cite{lyons2016guidance}. It acknowledges that e.g.~the choice to drive a car is not solely the result of people’s individual preferences, i.e. exogenous to the system, but determined largely by transport and urban systems organised around car driving, which can be changed with appropriate policies. 

Such policies can leverage long-term systems dynamics, allowing to re-shape transport systems for the better \cite{oecd2021tsn}. For example, the reversal of induced demand, namely disappearing traffic, happens when public space is reallocated from private cars to space-efficient modes of transport. As research shows \cite{cairns2002disappearing,itf2021rcp}, significant reductions in overall traffic levels can occur in the right conditions, because mode choice is more elastic than commonly believed: Many individuals are willing to give up traveling by car if there are appropriate alternatives. To succeed, it is also important to consider the right communication strategies and the strong psychological aspect of car attachment \cite{gossling2020cities}. Apart from policies to develop multimodal networks and street redesign, a large array of further policies can address a variety of system dynamics: From market-based instruments such as carbon-prices, road pricing, to abolishing minimum parking requirements, or financial support to increase the attractiveness of micromobility \cite{oecd2021tsn}. For all these policies, it is important to prioritize correctly, for example following an ``avoid-shift-improve'' approach \cite{bongardt2019sustainable,holden2020grand}: First, focus on avoiding the need for mobility, for example through planning for proximity such as the 15-minute city \cite{moreno2021introducing}. Second, aim for a shift from unsustainable to sustainable modes of transport. Only as a last third step aim to improve existing components of the system such as making vehicle technology more efficient.

\section{Tools \label{sec:tools}}

Over the years, computational tools have become more and more important for studying urban systems and human mobility. For an overview of the available tools in geographic analysis and transport planning see Ref.~\cite{lovelace2021open}. Here we focus on open-source tools as they allow fully reproducible research or analysis. There is a plethora of tools specialized on various purposes, including network analysis (single-layer or multilayer), routing and access (unimodal or multimodal), or mobility analysis. 
We classify these computational tools into three groups.

\subsection{Transport and mobility tools}
We first describe the most general tools for mobility and transport analysis.

To work with public transportation data, the Python library \textit{transitfeed} \cite{google2020gtfs} is suited to parse, validate and build GTFS files. This tool is particularly useful to those interested in the manipulation of the raw data. However, to convert the data into a network, some additional steps are needed, such as using \textit{Peartree} (see below).

\textit{Movingpandas} \cite{graser2019movingpandas} is a Python package that provides trajectory data structures and functions for the analysis and visualisation of mobility data. In a similar sense, and also developed in Python, \textit{scikit-mobility}~\cite{pappalardo2021scikitmobility} is a library that implements a framework for analyzing statistical patterns and modeling mobility, including functions for estimating movement between zones using spatial interaction models, and tools to asses privacy risks related to the analysis of mobility datasets.

\subsection{Transport network tools}
In this section we cover tools for studying transportation networks. 

Multiple tools were developed to obtain data on transportation and multimodal infrastructures. One of the best known is \textit{OSMnx} \cite{boeing2017osmnx}, a Python package that downloads street networks from OpenStreetMap into Python objects. \textit{OSMnx} can further be used to download other transportation networks, and build its multimodal transport networks. 

Another reliable Python library to read data from OpenStreetMap and extract transportation networks is \textit{Pyrosm} \cite{tenkanen2020pyrosm}. Differently from \textit{OSMnx}, \textit{Pyrosm} reads the data directly from OpenStreetMap's Protocol Buffer Format files (*.osm.pbf), while \textit{OSMnx} downloads the data from the Overpass API~\cite{overpass}. For this reason \textit{Pyrosm} is a particularly good alternative when working with large urban areas, states, and even countries, while \textit{OSMnx} typically offers a more precise way to collect data from specific points in a city.

Finally, an alternative to \textit{transitfeed}'s functionality of reading GTFS feeds is \textit{Peartree} \cite{butts2021peartree}, a Python library allowing to convert GTFS feed schedules into the corresponding directed network graph.

\subsection{Tools for routing or access on transport networks}
In this section we cover tools that can be used for routing and navigation. 

On one hand there are established general, high-performance  tools for unimodal routing such as \textit{graphhopper} \cite{graphhopper} and \textit{OSRM} \cite{osrm}. These tools were developed for routing on one network type but could in principle be extended to multimodal routing. Explicit multimodal routing is provided by \textit{OpenTripPlanner (OTP)} \cite{otp} and \textit{R5} \cite{conway2017evidence}. Both tools exist as fast R implementations, \textit{r5r} \cite{pereira2021r5r} and \textit{OTP for R} \cite{morgan2019opentripplanner}; the Python implementation of \textit{R5} is \textit{R5py} \cite{r5py}. The R package \textit{dodgr} \cite{padgham2019dodgr}, thanks to an efficient algorithm for computing distances between pairs of nodes in a street network, further enables the aggregation of flows on network links from a set of origin and destination points, a matrix of pairwise flow densities, and the underlying street network.

Finally, several open source packages focus on computing accessibility metrics, e.g.~the ease by which people can reach points of interests, such as those offering employment, shopping, medical care or recreation \cite{rietveld2000accessibility}. \textit{Pandana} \cite{foti2012generalized,pandana} is a Python library that enables to compute the accessibility of places by retrieving points of interest and street network data from OpenStreetMap~\cite{OpenStreetMap} and by efficiently computing shortest paths along the street network. \textit{Access} \cite{saxon2021open} is a Python library that computes a wide range of spatial accessibility metrics from a set of origins and destinations, and travel times or distances between them. Built on top of \textit{Pandana}, \textit{UrbanAccess} \cite{blanchard2017urbanaccess} integrates the creation of multi-modal transport networks (transit and pedestrian) using GTFS data and the computation of accessibility metrics. Similar functionalities are offered in R and Python by \textit{r5r} \cite{pereira2021r5r} and \textit{R5py} \cite{r5py}, respectively.

\section{Conclusions \label{sec:conclusion}}
In this chapter we gave a brief, introductory overview of quantitative urban mobility research from a complexity science perspective with hand-picked highlights. We covered history, statistical analysis of trajectories, the systemic perspective, and a collection of state of the art software tools. Our chapter could serve as a basis for teaching the topic to an advanced audience (master level or higher); much of the discussed material is covered in open teaching resources available at: \href{https://github.com/mszell/geospatialdatascience}{https://github.com/mszell/geospatialdatascience}.

\bibliographystyle{spphys}
\bibliography{main}
\end{document}